\documentclass[conference]{IEEEtran}
\IEEEoverridecommandlockouts
\usepackage{cite}
\usepackage{amsmath,amssymb,amsfonts}
\usepackage{algorithm}
\usepackage{algpseudocode}
\usepackage{graphicx}
\usepackage{textcomp}
\usepackage{xcolor}
\usepackage[multiple]{footmisc}
\usepackage{caption}
\usepackage{subcaption}
\usepackage{color}
\usepackage{eso-pic}
\usepackage{numprint}
\usepackage{balance}

\algnewcommand\algorithmicinput{\textbf{Input:}}
\algnewcommand\algorithmicoutput{\textbf{Output:}}
\algnewcommand\Input{\item[\algorithmicinput]}%
\algnewcommand\Output{\item[\algorithmicoutput]}%

\usepackage[pdftex]{hyperref}
\usepackage{tikz}

\newcommand\copyrighttext{%
    \footnotesize \textcopyright 2023 IEEE. Personal use of this material is permitted. Permission from IEEE must be obtained for all other uses, in any current or future media, including reprinting/republishing this material for advertising or promotional purposes, creating new collective works, for resale or redistribution to servers or lists, or reuse of any copyrighted component of this work in other works.
    DOI: \href{https://doi.org/10.1109/BigData59044.2023.10386837}{https://doi.org/10.1109/BigData59044.2023.10386837}}
\newcommand\copyrightnotice{%
    \begin{tikzpicture}[remember picture,overlay]
        \node[anchor=south,yshift=10pt] at (current page.south) {\fbox{\parbox{\dimexpr\textwidth-\fboxsep-\fboxrule\relax}{\copyrighttext}}};
    \end{tikzpicture}%
}

\def\BibTeX{{\rm B\kern-.05em{\sc i\kern-.025em b}\kern-.08em
    T\kern-.1667em\lower.7ex\hbox{E}\kern-.125emX}}

\begin{document}

\title{Predicting Dynamic Memory Requirements for Scientific Workflow Tasks}

\author{
    \IEEEauthorblockN{Jonathan Bader\IEEEauthorrefmark{1}\IEEEauthorrefmark{3}, Nils Diedrich\IEEEauthorrefmark{1}\IEEEauthorrefmark{3}, Lauritz Thamsen\IEEEauthorrefmark{4},
        and Odej Kao\IEEEauthorrefmark{3}}

    \IEEEauthorblockA{
        \IEEEauthorrefmark{3}
        \{firstname.lastname\}@tu-berlin.de, Technische Universität Berlin, Germany\\
    }
    \IEEEauthorblockA{
        \IEEEauthorrefmark{4}
        lauritz.thamsen@glasgow.ac.uk, University of Glasgow, United Kingdom\\
    }

}
\IEEEpubid{\makebox[\columnwidth]{*equal contribution \hfill} \hspace{\columnsep}\makebox[\columnwidth]{ }}

\maketitle
\copyrightnotice
\pagestyle{plain}

\begin{abstract}
    With the increasing amount of data available to scientists in disciplines as diverse as bioinformatics, physics, and remote sensing, scientific workflow systems are becoming increasingly important for composing and executing scalable data analysis pipelines.
    When writing such workflows, users need to specify the resources to be reserved for tasks so that sufficient resources are allocated on the target cluster infrastructure.
    Crucially, underestimating a task's memory requirements can result in task failures. 
    Therefore, users often resort to overprovisioning, resulting in significant resource wastage and decreased throughput.

    In this paper, we propose a novel online method that uses monitoring time series data to predict task memory usage in order to reduce the memory wastage of scientific workflow tasks.
    Our method predicts a task's runtime, divides it into k equally-sized segments, and learns the peak memory value for each segment depending on the total file input size.
    We evaluate the prototype implementation of our method using workflows from the publicly available nf-core repository, showing an average memory wastage reduction of 29.48\% compared to the best state-of-the-art approach.

\end{abstract}

\begin{IEEEkeywords}
    Resource Management, Scientific Workflow, Memory Prediction, Cluster Computing, Machine Learning
\end{IEEEkeywords}

\section{Introduction}\label{sec:INTRO}
Recent years have shown an increasing amount of generated data in all fields of science.
For instance, the Square Kilometre Array (SKA) telescopes generate terabytes of raw data per second~\cite{farnes2018building} and the Large Hadron Collider (LHC) retains petabytes of data per year~\cite{barisits2018atlas}.
Due to the increasing amount of data, the manual handling of tasks as a sequence of scripts is no longer feasible, and running them on a single scientist's personal computer is impractical. 
As a result, workflows are frequently employed to automate, parallelize, and monitor the data analysis process~\cite{da2023workflows,bader2022towards,deelman2018future}.

Such workflows consist of a set of tasks that act as a wrapper for an arbitrary application and transform data inputs into data outputs.
The tasks' execution order is defined by their interdependencies, defining the workflow's dataflow.
Due to the amount of data and the associated long runtimes, workflows are frequently executed on large cluster infrastructures administrated by resource managers such as Slurm~\cite{yoo2003slurm} or Kubernetes~\cite{burns2016borg}.
Resource managers rely on resource estimates, e.g., the peak memory or the number of cores, to allocate each task to a suitable machine~\cite{lehmannHowWorkflowEngines2023} and to ensure user resource limits.
Usually, these estimates are provided by the user and are, therefore, error-prone and inaccurate~\cite{phung2021not,hirales2012multiple,witt2019learning}.
Due to the risk of bottlenecked or even failed task executions caused by assigning too little memory~\cite{losser2022bottlemod,kintsakis2019reinforcement}, scientists often lean towards overprovisioning memory~\cite{tovar2022dynamic}.
This leads to a wastage of resources, reducing the cluster's throughput and increasing the cost~\cite{tovar2017job}.

\begin{figure}[t]
    \centering
    \includegraphics[width=0.9\columnwidth]{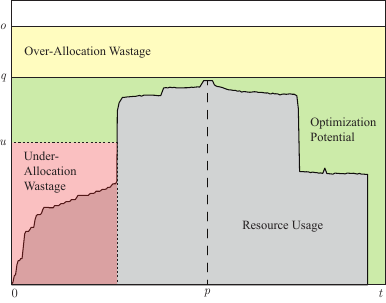}
    \caption{The figure shows a task's memory usage over time, the optimal, under-, and over-allocation when predicting a single peak memory value, as well as the associated optimization potential.}
	\label{fig:waste_overview}
\end{figure}

There are methods available for predicting a task's memory requirements without the need for scientists to provide manual estimates.
State-of-the-art methods approach this problem by predicting a task's peak memory with analytic methods~\cite{tovar2017job}, regression models~\cite{witt2019feedback,witt2019learning}, or reinforcement learning~\cite{bader2022leveraging}.
However, they neglect the fact that a task's resource consumption varies over time.
For instance, Figure~\ref{fig:waste_overview} shows a task's memory consumption function $f$ over time that peaks at point $p$ and flattens out afterward.
Allocating $q$ GB of memory leads to a successful task execution as $f(p)$ is the global maximum and $f(p) < q$.
However, as memory usage fluctuates, allocating a fixed amount of memory during a task's complete lifetime does not necessarily reflect the task's actual memory usage needs.
This observation indicates significant optimization potential currently overlooked by existing static memory estimation approaches.

Our paper addresses this by leveraging time series monitoring data to predict the memory consumption of scientific workflow tasks over time at a fine-grained level.
Our method employs a two-step approach, separating the dynamic memory prediction into distinct runtime and static memory prediction steps.
First, we predict a task's runtime with a linear regression model depending on the task's data input size.
To this prediction, we add a negative offset in order to prevent task failure due to a delayed increase of memory caused by an expected longer task runtime.
Second, we divide the time series into $k$ segments and train $k$ linear regression models, predicting the peak memory usage for each respective segment.
To safeguard task executions against prediction errors, we offset predictions with a buffer based on the observed prediction errors.

The contributions of this paper are:

\begin{itemize}
    \item We propose the k-Segments method that uses time series data to predict a task's memory usage over time. 
    To this end, we separately predict a task's runtime and peak memory values over $k$ segments, merging the results into a memory prediction function.
    \item We provide a prototype implementation of our method\footnote{\url{https://github.com/dos-group/k-Segments}} and publish our experimental traces, which showcase task resource usages\footnote{\url{https://github.com/dos-group/k-Segments-traces}}.
    \item We evaluate our k-Segments method with two real-world workflows consisting of 33 different tasks in total.
    Our experimental results show an average memory wastage reduction of 29.48\% compared to the best state-of-the-art method.
    
\end{itemize}

\emph{Outline}. The remainder of the paper is structured as follows.
Section~\ref{sec:RELATED_WORK} explains how state-of-the-art methods handle memory prediction for workflow tasks.
Section~\ref{sec:APPROACH} presents our time series-based memory prediction method.
Section~\ref{sec:EVALUATION} examines our method through a comparison with state-of-the-art baselines and an analysis of the results obtained.
Section~\ref{sec:CONCLUSION} summarizes and concludes our paper.

\section{Related Work}\label{sec:RELATED_WORK}
This section commences with covering related research on workflow task memory prediction. 
Subsequently, we provide a comparative analysis, highlighting distinctions between our method and previous approaches.

\subsection{Workflow Task Memory Prediction}

This subsection covers research on workflow task memory prediction.

Tovar et al.~\cite{tovar2017job} presented a task sizing strategy for high-throughput scientific workflows. 
In their work, they propose a model that predicts the peak resource usage for a specific task using an analytical model.
The model can be tweaked towards two different objectives: either maximizing throughput or minimizing wastage.
Both objectives optimize resource usage under the assumption of a slow-peaks model.
The slow-peaks model assumes a worst-case scenario where tasks fail at the end of their execution. 
The authors suggest using a two-step policy that initially allocates the predicted amount of resources, followed by the maximum available resources in case the first allocation results in an under-allocation. 
The authors shortly discuss that a multi-step policy is possible but needs further evaluation. 
Their results show that their approach achieves an overall increase in throughput and a decrease in resource wastage.

Witt et al.~\cite{witt2019feedback} used an online feedback loop-based resource allocation method to minimize resource wastage.
To learn the resource usage, their predictor uses the
sum of the task’s input files and trains different linear regression models for peak memory usage prediction. 
In contrast, Tovar et al.~\cite{tovar2017job} do not use a predictor and base their prediction only on the historical peak usages. 
Witt et al. dynamically offset the linear regression to achieve an overprediction and avoid task failure through underprovisioning. 
There are multiple offset strategies proposed.
The offset strategy LR mean ± adds the standard deviation as an offset, whereas the LR mean - approach only considers negative prediction errors.
The LR max offset strategy adds the largest observed underprediction as an offset. 
The results yield that their work improves resource utilization and is even able to outperform the work by Tovar et al.~\cite{tovar2017job}.

Addressing the memory allocation problem, Witt et al.~\cite{witt2019learning} propose a second method that minimizes the wastage and not the prediction error.
The authors assume a relationship between input data size and a task's resource usage and train a linear model based on this assumption.
They examine different failure handling strategies, i.e., how to address task failures due to underestimating memory.
In their implementation, they decide on a strategy that restarts the task with twice the amount of requested memory.
Their evaluation shows that selecting the appropriate failure-handling strategy has a considerable impact on wasted resources.

In our own related work~\cite{bader2022leveraging}, we proposed two different reinforcement learning approaches based on gradient bandits and Q-learning. 
Both methods have the objective of minimizing resource wastage.
Contrary to the previously discussed methods, the two reinforcement learning bandits did not implement an offset technique. 
We examined actions, state spaces, policies, and reward functions for the two reinforcement learning methods, showing how to circumvent limitations such as discrete action spaces.
The evaluation shows that out of the two proposed reinforcement learning methods, the gradient bandit performs better, outperforming the feedback loop-based approach by Witt et al.~\cite{witt2019feedback} and significantly reducing memory wastage compared to the default configuration.

\begin{figure*}[!ht]
	\centering
	\includegraphics[width=.95\textwidth]{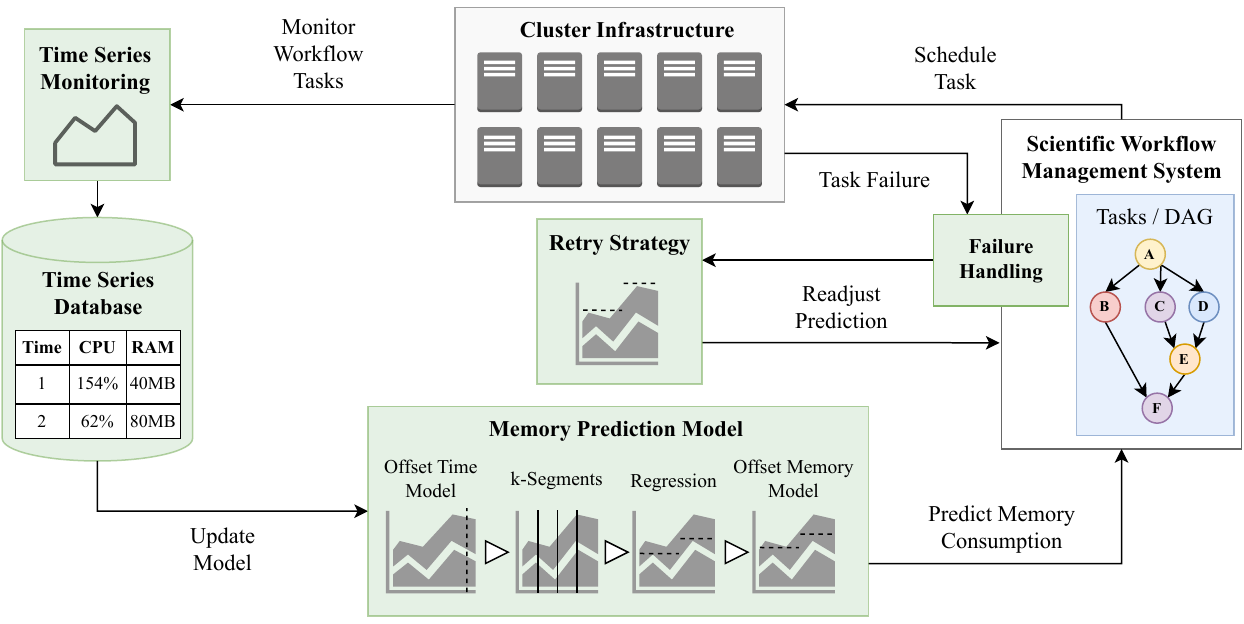}
  \caption{The figure provides a high-level overview of our k-Segments method (green) as applied in a scientific workflow environment. 
  The model of our method learns time-dependent memory allocations by underpredicting the runtime, dividing a task's time series into $k$ equally distributed segments, learning the peak memory value per segment, and offsetting it.
  The predictions are updated during the runtime of a workflow execution and provided to the workflow management engine.}
	\label{fig:approach_overview}
\end{figure*}

\subsection{Comparison with Own Method}

In this subsection, we compare the previously related work with our own k-Segments method.

The presented methods predict the peak memory consumption of a single task instance.
In contrast, our method predicts $k$ peaks over the time a task is executed.
Therefore, we use a new linear model for all $k$ segments in the time series.
This can potentially lead to lower memory wastage as a task's memory usage can be granularly defined but opens the space for more frequent task failures due to underprovisioning memory.
Thus, similar to the related work, we employ an offset strategy to avoid such failures.
However, we have to employ a more conservative strategy to avoid task failures, using the largest historical prediction error.
In addition, we need to consider offsetting only failed segments or each segment.

\section{Approach}\label{sec:APPROACH}

This section provides a comprehensive overview of our method, including a detailed description of its runtime and memory prediction components, their integrated application for dynamic memory consumption prediction, and our approach to addressing prediction failures.

\subsection{Overview}

Figure~\ref{fig:approach_overview} provides an overview of the execution environment in which our method is applied.
The scientific workflow management system (SWMS) is responsible for submitting the workflow tasks to the cluster infrastructure.
During the execution of the workflow tasks, our monitoring component collects time series monitoring data such as memory usage, CPU usage, or file events and stores it in a database.
On task completion, our method retrieves the monitoring data from the database and builds or updates its task resource prediction model.
The model partitions the time series into $k$ segments and trains a linear regression model for each segment, predicting the segment's memory usage.
Whenever the SWMS submits a known workflow task, it receives a predicted resource allocation function for the respective task from our model.
Our method works in an online fashion, meaning that the task resource prediction model is trained during the execution of the workflow.
This especially benefits workflow and task executions without historical data, as an offline method cannot provide resource predictions in such cases, leading to the use of user defaults.

\subsection{k-Segments Model Creation}

\begin{figure}[t]
	\centering
	\includegraphics[width=1.\columnwidth]{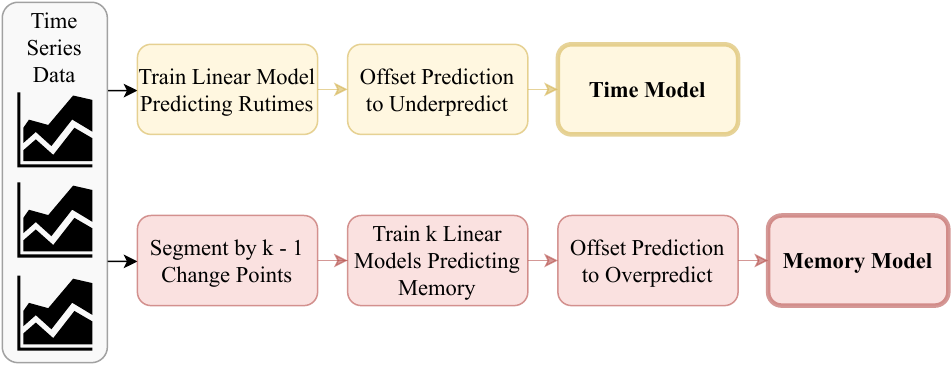}
  \caption{The figure shows the prediction model creation steps divided into runtime prediction and memory prediction}
	\label{fig:model_creation}
\end{figure}

Figure~\ref{fig:model_creation} shows the model creation steps, which we cover in the following in detail.
Contrary to the traditional approaches that predict peak memory usage, our time series-based method needs to take into account differences in task runtimes.
Therefore, as a first step, our model predicts the task's expected runtime.
We use a linear regression that assumes a relationship between the task's input size and its runtime.
Given a dataset $D = \{(x_1, Y_1),...,(x_n,Y_n)\} $ of a single task type, where $x$ is a scalar representing the total size of the input file and $Y$ is a list representing the memory usage over time of a single task execution.
Then, a linear model can be trained with the total input file size $x$ as the independent variable and the runtime $r$ as the dependent variable, where $r = j \cdot f$, $j = length(Y)$, and $f$ is equals to the length of the monitoring interval.
Then, we subtract the largest negative historical prediction error of our prediction as an offset.

Next, we define $k-1$ change points that are evenly distributed across the time series, creating $k$ segments in total.
Again, given a dataset $D = \{(x_1, Y_1),...,(x_n,Y_n)\}$ of a task type, where $x$ is a scalar, the total size of the input file, and $Y$ is a list of the memory usage over time of a single task execution.
Each $Y$ of size $j$, $j = length(Y)$, can be transformed into a segmented time series $Y^*$ with $k$ segments using $k-1$ change points in form of:
\begin{multline*}
Y^* = (s_{1},...,s_{k}) =((y_1,...,y_i),(y_{i+1},...,y_{2i}),...,\\(y_{(k-2)i+1},...,y_{(k-1)i}),(y_{(k-1)i +1},..., y_{j})) \text{, where $i = \lfloor\frac{j}{k}\rfloor$. }
\end{multline*}
Then, we have to provide a function that returns a segment's memory usage.
Therefore, we use the transformed dataset $D^* = \{(x_1, Y^*_1),...,(x_n,Y^*_n)\}$ and calculate the peak for each datapoint's segment, i.e., $Y^{**} = ( max(s_1),...,max(s_k))$.
This simplification enables us to train $k$ linear regressions, one for each segment, as the segment values are now scalar.
After training these regressions, we add the largest positive prediction error from historical executions on the regressions' intercepts to avoid underpredictions.

\subsection{Time Series Memory Prediction}
After the models are trained, they can be used to predict a resource allocation function for the next task execution.
First, we obtain a task's input data size from the SWMS and use it to train the time and the memory model.
Then, we predict the task's expected runtime $r_e$ with the time model and split it into $k$ values $R=( r_1 = r_s ,r_2 = 2r_s,...,r_{k-1} = (k-1)r_s, r_k = r_e)$, where $r_s = \lfloor\frac{r_e}{k}\rfloor$.    
The memory model predicts $k$ new resource allocations $V = (v_1,...,v_k)$ where $v>0$.
If a segment's predicted memory value is smaller than the previous segment's memory prediction, i.e., ${v_{k-1} > v_{k}}$, we take the previous segment's memory prediction.
For predictions where $v_1<0$, we set the default value.
This ensures that the function increases monotonically.
By combining the two sets of data, the following monotonically increasing step function can be constructed:

\begin{equation}
    \label{eq:step_function}
    f(x) =
    \begin{cases}
        v_1,& \text{if } 0 \leq x \leq r_1 \\
        v_2,& \text{if } r_1 < x \leq r_2 \\
        \hspace{5pt} \vdots & \\
        v_k,& \text{if } r_{k-1} < x \leq r_k \\
    \end{cases}
\end{equation}

\begin{figure}[t]
    \centering
    \includegraphics[width=0.9\columnwidth]{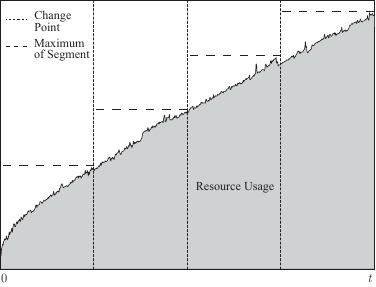}
    \caption{Example of applying our k-Segments method to the adapter removal task with $k = 4$.}
    \label{fig:k_segments_example}
\end{figure}

Figure~\ref{fig:k_segments_example} shows such a step function with 4 segments applied to the real-world adapter removal workflow task.

\subsection{Failure Handling}

\begin{figure}[t]
	\centering
	\includegraphics[width=0.9\columnwidth]{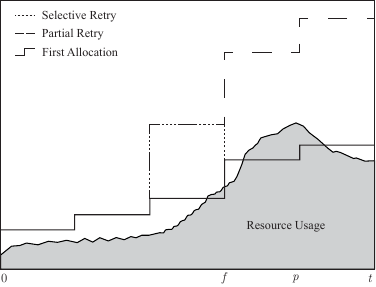}
  \caption{The figure illustrates the selective and partial retry strategies and their effect on a retried task execution.}
	\label{fig:retry_strategies}
\end{figure}

Allocating too little memory will cause the task to fail, and the task must be retried.
This can happen even if we use offsetting strategies.
Therefore, we propose two different failure-handling strategies. 
First, the \textbf{Selective Retry Strategy} that adjusts only the failed segment.
Second, the \textbf{Partial Retry Strategy} that adjusts every segment starting from the one that caused the failure.
When adjusting the segment(s), we have to select the new memory values.
Therefore, we define the retry factor $l$, which is multiplied by the failed allocation $v$ to provide the new allocation $v_{new}$.
Figure~\ref{fig:retry_strategies} shows the two different failing strategies with a retry factor of $l = 2$.
Here, a selective retry strategy would again lead to a task failure as the fourth segment still underpredicts the memory. 

In general, the efficacy of failure strategies depends on the task's memory consumption, and there is no overall superior strategy~\cite{phung2021not,witt2019learning}.
In our time series-based method, also the number of upcoming segments and their chance of failure affect the efficacy of the strategy.
For instance, if only one segment would cause failure, which we do not know in advance, the selective retry strategy would perform better.
Lastly, we would like to mention that there is also the possibility of adjusting a task's predicted runtime instead of the predicted memory, but this type of failure strategy is not considered in our work.

\section{Evaluation}\label{sec:EVALUATION}
Our evaluation section is divided into five key subsections: Prototype Implementation, Experimental Setup, Baseline Methods, Results, and Discussion. 
These components collectively provide a thorough assessment of our method's performance and its implications in the context of memory prediction for scientific workflow tasks.

\subsection{Prototype Implementation}

\begin{figure}[t]
	\centering
	\includegraphics[width=1.0\columnwidth]{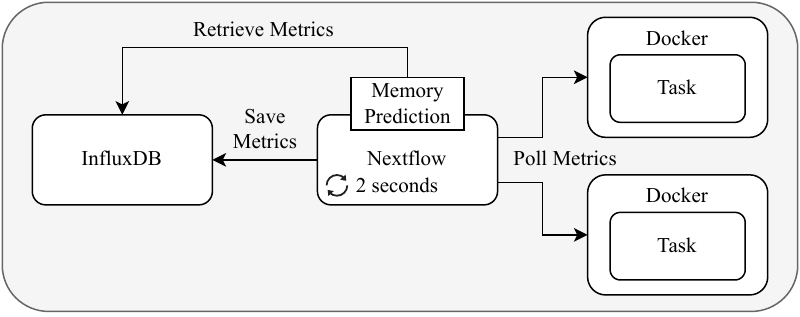}
  \caption{The figure provides an overview of our prototype implementation and its interaction with the existing workflow management system Nextflow.}
	\label{fig:implementation}
\end{figure}

Figure~\ref{fig:implementation} provides a simplified overview of our implementation.
We developed our method as an extension for the SWMS Nextflow~\cite{nextflow}.
By default, Nextflow collects peak and average resource usage values.
We extended this with a time series monitoring component that uses InfluxDB to store periodic metrics.
Our monitoring extension uses the Docker API to extract measurements.
Docker uses cgroups, a Linux kernel feature that allows processes to be organized in groups that can then be limited and monitored~\cite{al2017autonomic,zhuang2017taming}.
Our monitoring extension accesses the cpuacct, memory, and blkio controller from the Linux kernel via the Docker API to retrieve the monitoring information.
Because our method utilizes these low-level metrics, it is also portable to non-containerized setups.
The length of the monitoring interval can be adjusted and comes with a default of two seconds.
We observed that two seconds does not significantly impact the performance while still being able to provide accurate measurements.
Lowering the interval length involves the risk of overlooking memory peaks.
In addition to monitoring the cgroups, we also monitor the files, providing information such as the number of input files or total input size.
The gathered information is then retrieved from the database by our memory predictor.
The memory predictor uses Python with NumPy and scikit-learn to build its models and keeps them in memory.
As defaults for our method, we use $k=4$ for the number of segments, a retry factor of $l=2$, and 100MB as the minimum amount of memory to allocate in case the model predicts an allocation of less than zero.

Please note that Figure~\ref{fig:implementation}, the implementation, refers to a simplified setup on a single node.
On a cluster, each node would provide a monitoring service that could store the data directly in the database, making it available to Nextflow and the memory predictor.

\subsection{Experimental Setup}

For our experiments, we run two real-world workflows from the nf-core repository~\cite{ewels2020nf}, namely eager~\cite{yates2021reproducible} and sarek~\cite{hanssen2023scalable, garcia2020sarek}.
For the workflow inputs, we used data available from two studies. 
The eager workflow input data is from a study published in 2018 by de Barros Damgaard et al. on the population history of the Eurasian steppe~\cite{damgaard2018137}.
The input data for the sarek workflow is from a study published in 2022 by Harrod et al.~\cite{harrod2022genome}.
The sarek workflow ran for approximately one day and 5 hours, has 29 different tasks with an average runtime from 2 seconds to 1 hour, and an average peak memory usage from 10 MB to
about 23 GB.
The eager workflow ran for three days and 12 hours, contains 18 different tasks with an average runtime between 8 seconds and 4 hours, and a peak memory usage from 19MB to 14GB.
The sarek workflow contains up to 1512 executions of the same task, and the eager workflow up to 136 executions of the same task.
For gathering the metrics, we run the workflows with our prototype implementation on a machine consisting of an AMD EPYC 7282 16-Core Processor (32 Threads), 128GB DDR4 memory, and two 960GB SATA III SSDs. 

We use the gathered metrics to feed our simulation tool.
Through simulation parameters, the proportion of training/test data can be defined.  
We simulate an online approach where finished task executions can be incorporated into the learning process respective to real-world systems.
This is also an important aspect as we want to incorporate state-of-the-art baselines that can use online learning.

\subsection{Baselines}

\begin{figure*}[t]
\centering
\begin{subfigure}{.95\textwidth}
   \centering
  \includegraphics[width=1\textwidth]{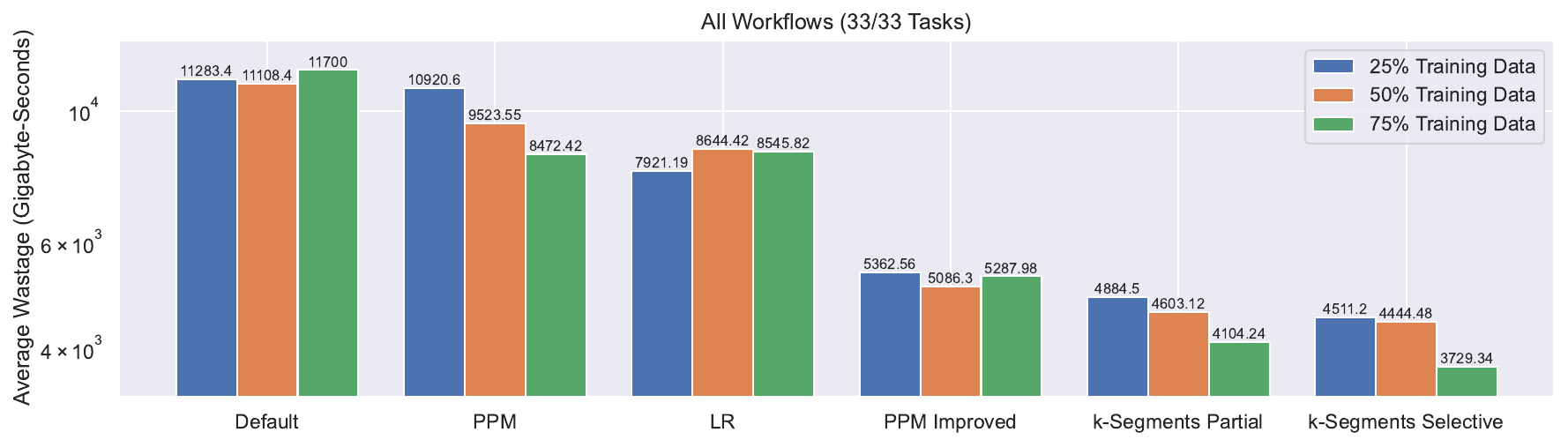}
  \caption{The figure compares the average wastage in GBs of our k-Segments-based methods to two state-of-the-art baselines, one state-of-the-art baseline improved by us, and the default configurations of the workflows across all 33 workflow tasks.}
  \label{fig:eval_wastage}
\end{subfigure}
\begin{subfigure}{.95\textwidth}
  \centering
  \includegraphics[width=1\textwidth]{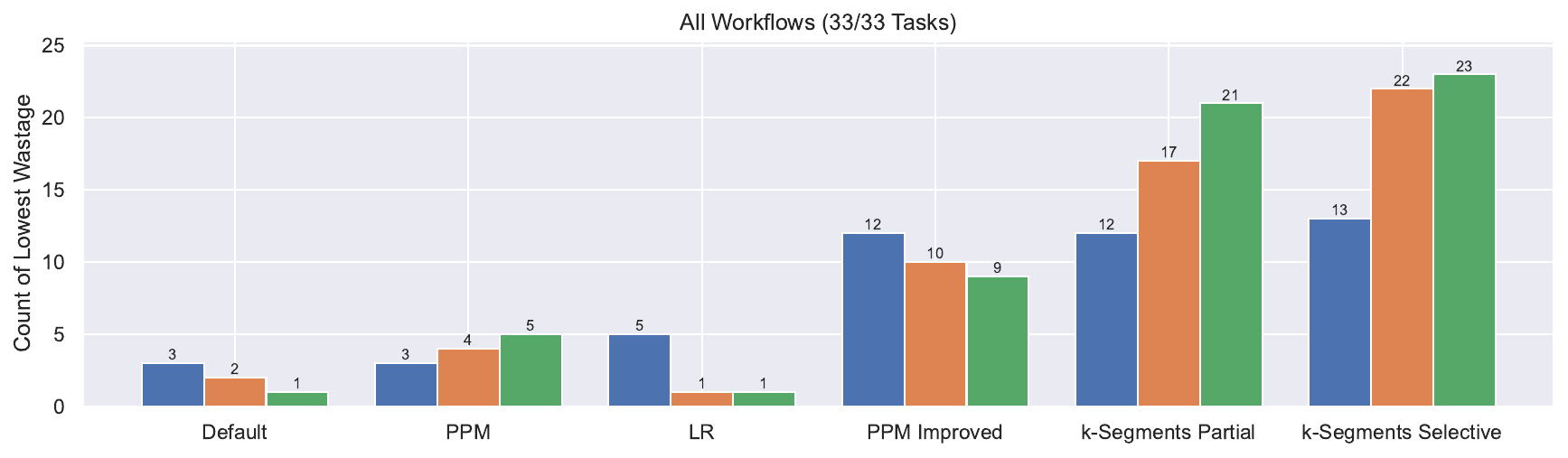}
  \caption{The figure compares the number of times a method achieves the lowest wastage across all 33 workflow tasks.}
  \label{fig:eval_counts}
\end{subfigure}
\begin{subfigure}{.95\textwidth}
   \centering
  \includegraphics[width=1\textwidth]{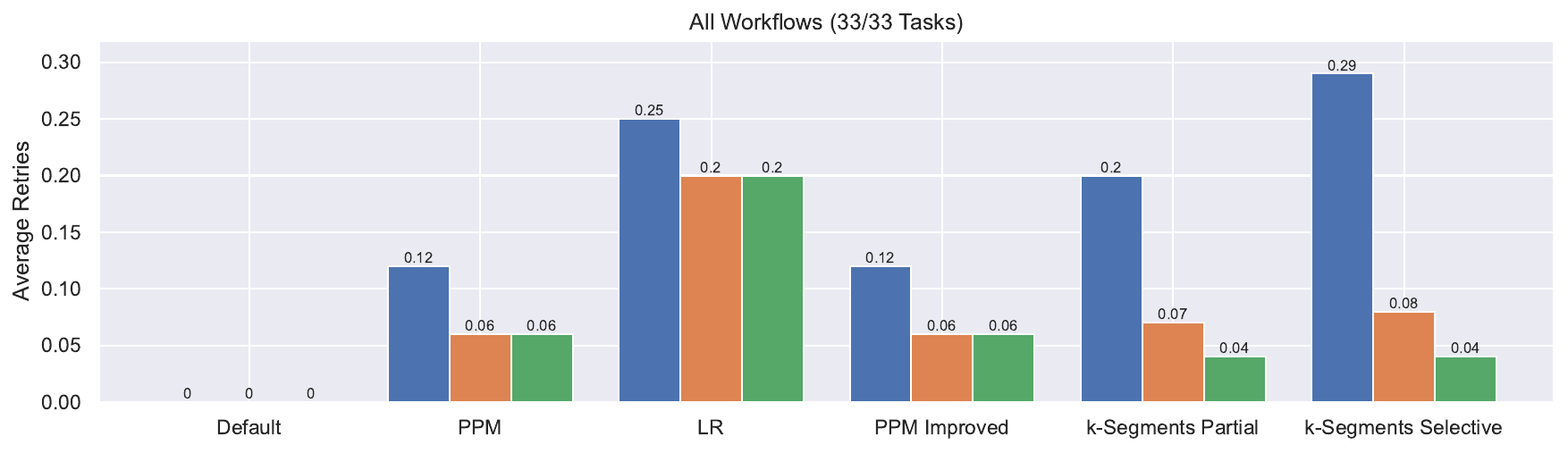}
  \caption{The figure compares the average retries of our k-Segments-based methods to two state-of-the-art baselines, one state-of-the-art baseline improved by us, and the default configurations of the workflows across all 33 workflow tasks.}
  \label{fig:eval_retries}
\end{subfigure}
\caption{The figures show the experimental results.}
\label{fig:eval_total}
\end{figure*}%

We compare our method with two state-of-the-art baselines, one state-of-the-art baseline that we have improved, and the default configurations of the workflows.

The workflows' default configurations are the ones provided by the workflow developers.
They are used when running the workflows out of the box and, therefore, serve as a sanity baseline.

Tovar et al.~\cite{tovar2017job} use probabilities of peak memory values (PPM) from historical executions to select a task's initial amount of peak memory.
Their method aims to minimize the sum of probabilities of resource peaks where the resource peak is greater than the allocated resource value.
If the first resource allocation leads to an underprediction and the task fails, Tovar et al. assign a node's maximum amount of memory.
We use the source code provided by the authors and integrate it into our simulation.
As an additional baseline, we extend the method of Tovar et al. and implement a failure strategy that doubles the memory upon failure instead of assigning a node's maximum memory immediately.
We name this baseline PPM Improved.

Witt et al.~\cite{witt2019feedback} use an online learning method that uses a linear regression model (LR) to estimate a task's peak memory, assuming a linear relationship between data input size and memory usage.
As an offset, they add the standard deviation based on historical predictions.
The authors assign double the amount of memory for failed tasks and execute them again.
Due to the lack of provided links to the source code, we had to implement the method according to the descriptions in the paper.

\subsection{Results}

Figure~\ref{fig:eval_total} shows the results of the memory prediction with three different amounts of training data for 33 workflow tasks from the two workflows.

Figure~\ref{fig:eval_wastage} presents the wastage of all approaches in gigabyte-seconds. 
Among these, the default baseline exhibits the highest wastage, approximately 2.5 to 3 times higher than the best-performing method, k-Segments Selective. 
Comparatively, the literature baselines manage to outperform the default but still demonstrate significantly higher wastage compared to our k-Segments methods.
Across all levels of training data, k-Segments Selective consistently achieves the lowest wastage, closely followed by k-Segments Partial. 
The k-Segments models demonstrate a wastage reduction of 22.39\% for the partial retry strategy and 29.48\% for the selective retry strategy compared to the best-performing baseline, PPM Improved, using 75\% of the training data.
Noticeably, while our methods, k-Segments Selective and k-Segments Partial, exhibit lower wastage with increasing training data, the same isn't necessarily true for the baselines. 
Some baselines exhibit slight performance degradation when provided with more data.

Figure~\ref{fig:eval_counts} shows the number of times a method scores the lowest wastage.
If two methods both have the least wastage, they both get one point.
Using 25\% of the training data, the k-Segments Selective method has the highest count, closely followed by k-Segments Partial and PPM Improved by Tovar et al., which yield identical counts.
For 25\% training data, PPM and the default baseline exhibit the lowest values. For 50\% and 75\% training data, LR by Witt et al. and the default baseline similarly display the lowest values.
Increasing the amount of training data for PPM Improved and the k-Segments methods shows an increase in counts for the k-Segments methods and a decrease for PPM Improved, again indicating that more training data positively affects our k-Segments methods.

Figure~\ref{fig:eval_retries} shows the methods' average number of retries.
The default strategy has an average of zero, indicating no task failures due to insufficient memory.
Using 25\% of the training data, our k-Segments Selective method has the highest number of failures, followed by the LR and the k-Segments Partial method.
PPM (Improved) shows the lowest average number of failures for a prediction method.
By increasing the training data to 50\%, PPM (Improved) and LR can reduce the average number of failures but show no improvement when providing more than 50\% of the training data.
Our k-Segments models are able to significantly reduce the average number of retries, even showing the lowest number achieved by any prediction method when using 75\% of the data.

\subsection{Discussion}

The evaluation results show that using our method with time series data to predict a task's memory usage leads to lower wastage.
Additionally, our method benefits from a growing quantity of training data, whereas the state-of-the-art methods do not necessarily exhibit improvements.
The increase in training data positively influences our method's retry count, thereby accounting for the lower wastage.

An unexpected outcome of the results is the substantial reduction in memory wastage observed using PPM Improved compared to its original implementation. 
On average, PPM Improved significantly outperforms Witt et al.'s LR method. 
We attribute this discrepancy to the improved failure-handling strategy, which was not assessed in Tovar et al.'s original paper but appears to yield superior outcomes. 
Due to the machines in our experimental setup being equipped with 128GB of memory, directly assigning the maximum amount of memory to a task upon failure leads to significant memory wastage.

\begin{figure}
\centering
\begin{subfigure}{1.\columnwidth}
   \centering
  \includegraphics[width=1\columnwidth]{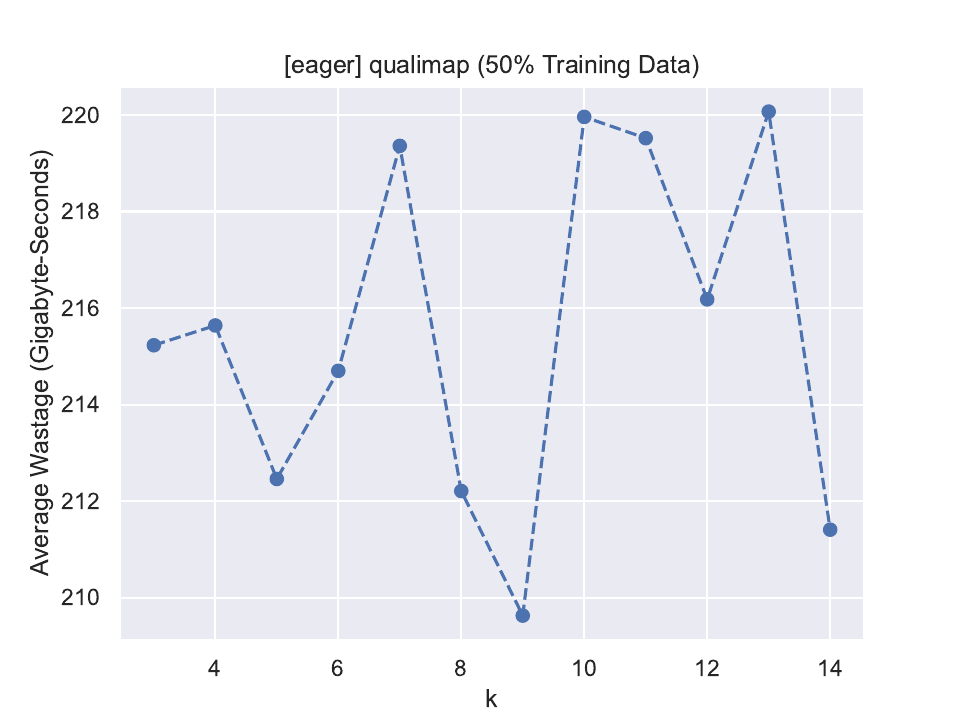}
  \caption{Qualimap task}
  \label{fig:discussion_qualimap}
\end{subfigure}
\begin{subfigure}{1.\columnwidth}
  \centering
  \includegraphics[width=1\columnwidth]{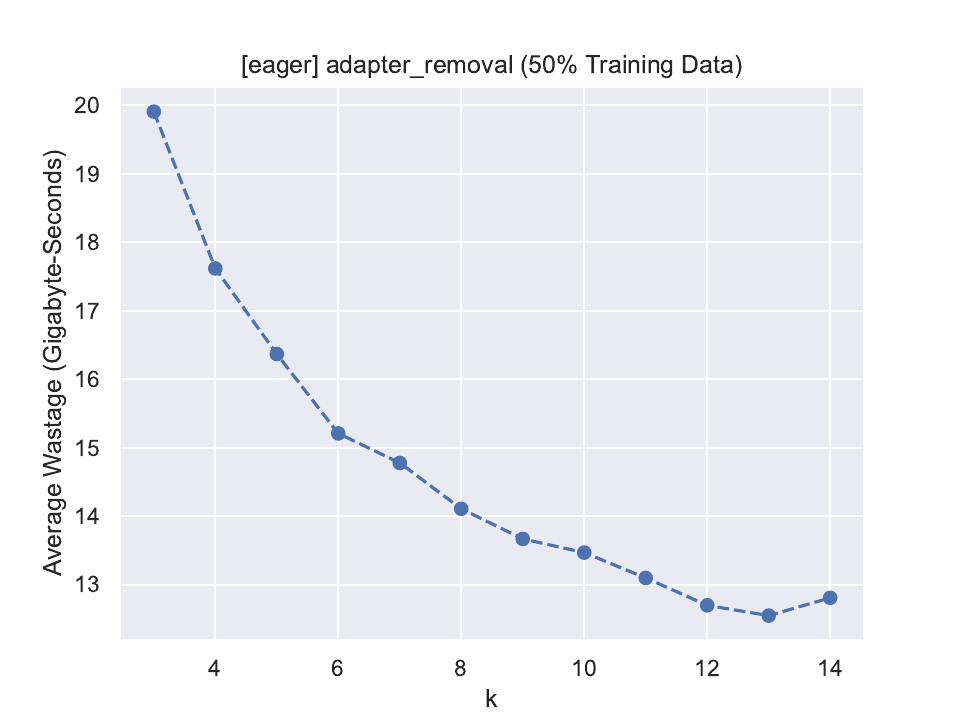}
  \caption{Adapter removal task}
  \label{fig:discussion_k_adapter}
\end{subfigure}
\caption{The figures show the memory wastage of the tasks using our method as a function of the chosen parameter $k$.} 
\label{fig:discussion_k}
\end{figure}

One of our method's limitations is the selection of the parameters, most importantly $k$.
In general, selecting a bigger $k$ can decrease the wastage but leads to possibly more error-prone assignments due to prediction errors that eventually increase the wastage.
For our experiments, we tested several values for $k$ and chose  $k=4$ for all tasks.
However, $k$ can also be selected for each task individually, posing a local optimization problem.
For instance, Figure~\ref{fig:discussion_k} presents two tasks from the eager workflow and the average wastage in relation to the chosen $k$ when using 50\% of training data.
The Qualimap task has a zigzag pattern with multiple local optima and a global optimum at $k=9$.
In contrast, the Adapter Removal task reduces its wastage up to $k=13$, making it easier to search for a global optimum.
As zigzag patterns such as in Figure~\ref{fig:discussion_qualimap} make it hard to deploy gradient-based search methods, reoptimizing $k$ on each iteration during online learning appears to be an option for this~\cite{witt2019learning}.

A second limitation of our method is its applicability in present-day systems.
Resource managers now require a memory estimate for the entire duration of a job's runtime while we provide an estimate over time. 
This necessitates that the allocated resources can be modified over time to derive benefits from the fine-grained predictions.
Additionally, scheduling must account for dynamic changes to memory over time.

\section{Conclusion}\label{sec:CONCLUSION}
In this paper, we presented our novel k-Segments method that predicts a workflow task's memory usage over time. 
To this end, our method predicts a task's runtime and divides the predicted time into $k$ equally-sized segments.
Then, our method predicts each segment's memory consumption depending on the task's input data size.
Using prediction offsets, the k-Segments model aims to avoid underallocations during the prediction process.

In our evaluation with two real-world scientific workflows from the publicly available nf-core repository, we show that our k-Segments method is able to outperform two state-of-the-art baselines, achieving a memory wastage reduction of 29.48\% compared to the best competitor.
Furthermore, our experiments illustrate that our method benefits significantly from an increasing amount of training data, with less wastage and fewer failures, compared to the state-of-the-art baselines.

Our method requires the number of segments $k$ as an input, which can vary between tasks.
In the future, we aim to explore methods of finding $k$ by using exploration and exploitation techniques from the field of reinforcement learning.
Additionally, our method's predictions are dynamic, requiring resource managers to support adjustments during runtime.
As this is not the current norm, we want to extend an existing resource manager's interface to cope with dynamic adjustments of memory claims.

\section*{Acknowledgments}
\thanks{Funded by the Deutsche Forschungsgemeinschaft (DFG, German Research Foundation) as FONDA (Project 414984028, SFB 1404).}
\balance
\bibliographystyle{IEEEtran}
\bibliography{./references}

\end{document}